\documentclass[%
 aip,
 amsmath,amssymb,
 reprint,%
]{revtex4-1}

\usepackage{graphicx}
\usepackage{dcolumn}
\usepackage{bm}

\usepackage[utf8]{inputenc}
\usepackage[T1]{fontenc}
\usepackage{mathptmx}
\usepackage{xcolor}

\usepackage{mathtools}
\usepackage{hyperref}
\DeclarePairedDelimiter\bra{\langle}{\rvert}
\DeclarePairedDelimiter\ket{\lvert}{\rangle}
\DeclarePairedDelimiterX\braket[2]{\langle}{\rangle}{#1 \delimsize\vert #2}

\begin{document}

\preprint{AIP/123-QED}

\title[First principles correction to core level LR-TDDFT]{First principles correction scheme for linear-response time-dependent density functional theory calculations of core electronic states}

\author{Augustin Bussy}%
 \email{augustin.bussy@chem.uzh.ch}
\altaffiliation{Department of Chemistry, University of Zurich, Winterthurerstrasse 190, CH-8057 Zürich, Switzerland.}%
\author{J\"urg Hutter}%
 \email{hutter@chem.uzh.ch}
\affiliation{Department of Chemistry, University of Zurich, Winterthurerstrasse 190, CH-8057 Zürich, Switzerland.
}%

\date{\today}

\begin{abstract}

Linear-response time-dependent density functional theory (LR-TDDFT) for core level spectroscopy using standard local functionals suffers from self-interaction error and a lack of orbital relaxation upon creation of the core hole. As a result, LR-TDDFT calculated X-ray absorption near edge structure (XANES) spectra need to be shifted along the energy axis to match experimental data. We propose a correction scheme based on many body perturbation theory to calculate the shift from first principles. The ionization potential of the core donor state is first computed and then substituted for the corresponding Kohn--Sham orbital energy, thus emulating Koopmans' condition. Both self-interaction error and orbital relaxation are taken into account. The method exploits the localized nature of core states for efficiency and integrates seamlessly in our previous implementation of core level LR-TDDFT, yielding corrected spectra in a single calculation. We benchmark the correction scheme on molecules at the K- and L-edges as well as for core binding energies and report accuracies comparable to higher order methods. We also demonstrate applicability in large and extended systems and discuss efficient approximations.

\end{abstract}

\maketitle

\section{Introduction}  

X-ray absorption spectroscopy (XAS) is a major characterization tool used in many fields of natural science. The technique provides a local and element specific probe, yielding insights into the geometric and electronic structure of matter. In particular, the X-ray absorption near edge structure (XANES) region of the spectrum holds information about the chemical state (coordination number, oxidation state, etc. ) of the probed atom. As technology progresses and quality light sources become more accessible, XANES is used increasingly often. Following this trend, many theoretical approaches have been developed to help interpret experiments.

One of the most wide spread computational method for the simulation of XANES is time-dependent density functional theory (TDDFT). It offers a favorable trade-off between cost and accuracy and is relatively easy to use due to its mostly "black-box" nature. In its standard formulation\cite{casida1995time, yabana1996time}, TDDFT is best suited for UV-Vis spectroscopy, where electronic transitions from valence to low lying unoccupied bound states take place. Much effort has been made in adapting the theory to core state spectroscopy. Most notably, core-valence separation\cite{cederbaum1980many} (CVS) has been used with both flavors of TDDFT, real-time\cite{lopata2012linear} (RT-TDDFT) and linear-repsonse\cite{stener2003time, besley2007time, george2008time} (LR-TDDFT), allowing the calculation of excitations from core states at an affordable cost. Other approaches rely on iterative solvers which can directly target high energy core transitions\cite{liang2011energy, schmidt2010assignment}, thus bypassing the CVS. The different implementations yield consistent results and the choice of the exchange correlation functional typically has a larger impact on accuracy\cite{besley2021modeling}.

While TDDFT generated XANES spectra are known to have accurate relative feature intensities and spacing, they usually have to be translated along the energy axis to match experiments. The main reasons behind this shortcoming of TDDFT lie in the lack of orbital relaxation upon the creation of a core hole and self-interaction errors\cite{imamura2007analysis} (SIE). There are multiple ways of dealing with these issues. The simplest is to apply a system dependent empirical shift\cite{minasian2014new, nardelli2011theoretical, li2016experimental}, although this limits TDDFT to a purely descriptive role. Similarly, an atom specific shift can be calibrated over multiple systems to be later applied to similar calculations\cite{debeer2010calibration, martin2015k}. A common approach to computing the shift from first principle is by performing a separate $\Delta$SCF\cite{norman2018simulating, besley2021modeling} calculation. By taking the difference between the converged ground and first excited states total energies, SIE mostly cancels out and orbital relaxation is taken into account. However, such calculations may be hard to setup and converge. Other techniques have been proposed to tackle SIE in TDDFT; for example by basing the TDDFT calculation on a SIE corrected ground state calculation\cite{tu2007self, imamura2007analysis}, fitting the fraction of Hartree--Fock exchange in hybrid functionals\cite{besley2007time} or developing core TDDFT specific functionals\cite{nakata2006hybrid, song2008core, besley2009time} (usually relying on empirically fitted parameters). For calculations involving excitations from heavy atoms, scalar relativistic effects do also play a role. This is usually taken into account at the ground state level, using the ZORA\cite{lenthe1993relativistic} or DKH\cite{hess1986relativistic} approach.

In this work, we propose a fully \textit{ab-initio} correction scheme for core LR-TDDFT spectroscopy based on many body perturbation theory. The method exploits the localized nature of core states for efficiency and seamlessly integrates into our previous implementation of core LR-TDDFT\cite{bussy2021efficient} in the CP2K software package\cite{kuhne2020cp2k}. The method accounts for orbital relaxation and self-interaction error, while yielding the core ionization potential as a side product. We discuss the method's implementation and demonstrate its applicability to molecules at the K- and L-edges and as a mean of calculating core binding energies. We also discuss efficient approximations and applications in extended systems.

\section{Theory}\label{theory}

The LR-TDDFT equations are built on the solutions of the time-independent Kohn--Sham (KS) equations. Expanded in a basis of atom-centered functions, the KS orbitals read

\begin{equation}
    \phi_{i\sigma}(\mathbf{r}) = \sum_p c^0_{pi\sigma}\varphi_p(\mathbf{r})
\end{equation}

\noindent where the basis elements $\{\varphi_p(\mathbf{r})\}$, also referred to as atomic orbitals (AOs), are typically non-orthogonal Gaussian type orbitals (GTOs) with

\begin{equation}\label{overlap}
    S_{pq} = \braket{\varphi_p}{\varphi_q}.
\end{equation}

\noindent Note that throughout this work, indices $p,q,...$ always refer to AOs, $i,j,...$ to occupied MOs, $a,b,...$ to virtual MOs and $\sigma, \tau, ...$ to spin.

In the Sternheimer\cite{sternheimer1951nuclear, hutter2003excited} approach to LR-TDDFT, vertical excitation energies are obtained by solving the non-Hermitian eigenvalue equation\cite{bussy2021efficient}:

\begin{equation}
    \label{full_prob}
    \omega \begin{pmatrix} -\mathbf{G} & 0 \\ 0 & \mathbf{G} \end{pmatrix} \begin{pmatrix} \mathbf{c}^+ \\ \mathbf{c}^- \end{pmatrix} = \begin{pmatrix} \mathbf{A+B-D} & \mathbf{B-E} \\ \mathbf{B-E} & \mathbf{A+B-D} \end{pmatrix} \begin{pmatrix} \mathbf{c}^+  \\ \mathbf{c}^-\end{pmatrix}
\end{equation}

\noindent where the $\mathbf{c}^\pm$ eigenvectors are the coefficients for the basis expansion of the LR orbitals and $\omega$ the corresponding excitation energy. The matrix $\mathbf{G}$ is related to basis set overlap:

\begin{equation}\label{metric}
    G_{pi\sigma,qj\tau} = S_{pq}\delta_{ij}\delta_{\sigma\tau},
\end{equation}

\noindent matrix $\mathbf{A}$ is based on the ground state KS matrix $\mathbf{F}^\sigma$:

\begin{equation}\label{matrix_A}
    A_{pi\sigma,qj\tau} = \left(F_{pq}^\sigma - \varepsilon_{i\sigma}S_{pq}\right)\delta_{ij}\delta_{\sigma\tau},
\end{equation}

\noindent where $\varepsilon_{i\sigma}$ is a KS eigenvalue, and $\mathbf{B}$, $\mathbf{D}$ and $\mathbf{E}$ are the Hartree exchange-correlation kernel, on- and off-diagonal Hartree--Fock (HF) exchange kernel matrices:

\begin{equation}\label{matrix_B}
    B_{pi\sigma, qj\tau} = \sum_{rs} Q^\sigma_{pr}(ri_\sigma|f^{Hxc}_{\sigma,\tau}|sj_\tau)Q^\tau_{qs},
\end{equation}

\begin{equation}\label{on-diag_ex}
    D_{pi\sigma, qj\tau} = c_{\text{HF}}\ \delta_{\sigma\tau}\sum_{rs} Q^\sigma_{pr}(rs|i_\sigma j_\tau)Q^\tau_{qs}
\end{equation}

\begin{equation}\label{off-diag_ex}
    E_{pi\sigma, qj\tau} = c_{\text{HF}}\ \delta_{\sigma\tau}\sum_{rs} Q^\sigma_{pr}(rj_\tau|si_\sigma )Q^\tau_{qs}
\end{equation}

\noindent where $\mathbf{Q^\sigma}$ is a projector on the unoccupied unperturbed space, $C_\text{HF}$ is the fraction of HF exchange and $f^{Hxc}_{\sigma,\tau}$ is defined as

\begin{equation}
    f^{Hxc}_{\sigma,\tau}(\mathbf{r},\mathbf{r'}) = \frac{1}{|\mathbf{r}-\mathbf{r'}|} + \left.\frac{\delta^2E_{xc}}{\delta n_\sigma(\mathbf{r}) \delta n_\tau(\mathbf{r'})}\right|_{n^0}
\end{equation}

\noindent within the adiabatic approximation\cite{gross1990time}.

The eigenvalue equation can be greatly simplified by setting $\mathbf{c}^+=0$ while retaining eigenvalue accuracy\cite{hirata1999time}. This reduces the matrix dimensions by a factor 2 and allows ignoring the off-diagonal exact exchange kernel matrix $\mathbf{E}$. This is known as the Tamm-Dancoff approximation\cite{fetter2012quantum} (TDA) and is consistently applied throughout this work. 

When dealing with core-spectroscopy, specific approximations can be made for efficiency. Firstly, core and valence excitations can be effectively decoupled, allowing the neglect of the latter\cite{cederbaum1980many}. This reduces the number of electron repulsion integrals (ERIs) to be evaluated as well as the dimension of the eigenvalue problem in equation (\ref{full_prob}). Moreover, core states of interests can be treated serially within the sudden approximation\cite{rehr1978extended, george2008time}, further reducing matrix dimensions to that of the ground state KS matrix. In this context, all the 4-center ERIs needed for the kernel matrices have either the form $(pI|qJ)$ or $(pq|IJ)$, where  $\phi_I$, $\phi_J$ are core donor MOs. In our recent implementation\cite{bussy2021efficient} of core level LR-TDDFT in the CP2K software package\cite{kuhne2020cp2k}, we introduced a local resolution of the identity (RI) scheme that further reduces the cost of ERIs:

\begin{equation}
    (pI|qJ) \approx \sum_{\mu,\nu}\ (pI|\mu) \ (\mu|\nu)^{-1} \ (\nu|qJ)
\end{equation}

and

\begin{equation}
    (pq|IJ) \approx \sum_{\mu, \nu}\ (pq|\mu) \ (\mu|\nu)^{-1} \ (\nu|IJ)
\end{equation}

\noindent where the RI basis $\{\chi_\mu(\mathbf{r})\}$ only consists of Gaussian functions centered on the excited atom. Such a choice of basis is only possible due to the localized nature of core states. In case of K-edge spectroscopy, only the 1s core MO is considered. For the L\textsubscript{2,3}-edge, the core MOs span all three 2p states. 

\begin{table*}
\caption{\label{table:k_edge} Comparison of GW2X corrected LR-TDDFT for XANES K-edge spectroscopy with similar methods from the literature. The difference between the calculated and experimental first allowed excitation is reported (in eV). The bold and italic atom in the chemical formula is the one from which the excitation takes place.}
\begin{ruledtabular}
\begin{tabular}{l c c c c c c c c c}
 & B3LYP-$\Delta$SCF\cite{besley2009self} & SCAN-ROKS\cite{hait2020highly} & SRC2-TDDFT\cite{besley2009time} & CIS(D)\cite{asmuruf2008calculation} & EOM-CCSD\cite{vidal2019new} & PBEh-GW2X & Exp. \\ 
 & \small{u6-311(2+,2+)G**} & \small{aug-cc-pCVTZ} & \small{6-311(2+,2+)G**} & \small{aug-cc-pCVQZ} & \small{aug-cc-pCVTZ \footnote{with additional Rydberg functions}} & \small{aug-pcX-2}& \\
\hline 
\rule{0pt}{2.6ex}\textit{\textbf{C}}H\textsubscript{4} &  +0.5 &  0.0 & --- &  + 0.3 & --- &  + 0.9 & 288.0\cite{schirmer1993k}\vspace{0.1cm}\\ 
\textit{\textbf{C}}\textsubscript{2}H\textsubscript{4} & --- &  0.0 &  + 0.6 & --- &  + 1.8 &  + 1.2 & 284.7\cite{hitchcock1977carbon}\vspace{0.1cm}\\ 
\textit{\textbf{C}}O &  - 0.8 &  - 0.4 &  - 0.7 &  + 2.5 &  - 0.4 &  + 2.5 & 287.4\cite{domke1990carbon}\vspace{0.1cm}\\ 
\textit{\textbf{C}}H\textsubscript{2}O &  - 0.1 & --- &  0.0 &  + 2.8 & --- &  + 1.7 & 286.0\cite{remmers1992high}\vspace{0.1cm}\\ 
\textit{\textbf{N}}H\textsubscript{3} & --- &  - 0.5 & --- & --- &  + 0.1 &  + 0.9 & 400.8\cite{schirmer1993k}\vspace{0.1cm}\\ 
\textit{\textbf{N}}NO &  - 0.7 &  - 0.1 & --- &  + 2.4 & --- &  + 0.2 & 401.0\cite{prince1999vibrational}\vspace{0.1cm}\\ 
N\textit{\textbf{N}}O &  - 0.8 &  - 0.2 & --- &  + 2.8 & --- &  + 0.4 & 404.6\cite{prince1999vibrational}\vspace{0.1cm}\\ 
C\textit{\textbf{O}} &  - 0.6 &  - 0.3 &  0.0 &  - 0.5 &  + 0.4 &  - 0.7 & 534.2\cite{domke1990carbon}\vspace{0.1cm}\\ 
H\textsubscript{2}\textit{\textbf{O}} & --- & --- & --- & --- &  + 0.4 &  -0.1 & 534.0\cite{schirmer1993k}\vspace{0.1cm}\\ 
CH\textsubscript{2}\textit{\textbf{O}} &  - 0.4 & --- &  0.0 &  + 1.0 & --- &  - 1.1 & 530.8\cite{remmers1992high}\vspace{0.1cm}\\ 
H\textit{\textbf{F}} & --- &  - 0.3 &  - 0.5 & --- &  + 0.4 &  - 0.8 & 687.4\cite{hitchcock1981k}\vspace{0.1cm}\\ 
\textit{\textbf{F}}\textsubscript{2} & --- &  - 0.2 & --- & --- & --- &  - 0.7 & 682.2\cite{hitchcock1981k}\vspace{0.1cm}\\ 
\hline 
\rule{0pt}{2.6ex}\textbf{MAD} & 0.6& 0.2& 0.3& 1.8& 0.6& 0.9& \\ 
\end{tabular}
\end{ruledtabular}

\end{table*}

LR-TDDFT yields excitation energies as corrections to ground state orbital energy differences\cite{ullrich2011time}, \textit{i.e.} 

\begin{equation} \label{tddft_simple}
    \omega = \varepsilon_a - \varepsilon_I + \Delta_{xc}
\end{equation}

\noindent where $\varepsilon_a$ is the orbital energy of a virtual receiving MO and $\varepsilon_I$ that of the core donor MO. Under Koopmans' condition, $\varepsilon_a$ and $\varepsilon_I$ are identified as the electron affinity (EA) and the negative ionization potential (IP), respectively. However, mostly because of self-interaction error, DFT orbital energies are far from the actual EAs and IPs and the condition does not hold\cite{dabo2010koopmans}. Nonetheless, and especially since $|\varepsilon_I| \gg |\varepsilon_a|$, LR-TDDFT core excitation energies are expected to be greatly improved if $-\varepsilon_I$ were to be substituted by an accurate value of the core IP in equation (\ref{matrix_A}).

Based on a Hartree-Fock ground state calculation, ionization potentials can be computed using second order electron propagator theory\cite{cederbaum1973direct, ortiz2013electron} by solving the following equation:

\begin{equation}\label{soep}
\begin{aligned}
     \text{IP}_I = - \varepsilon_I \ -& \frac{1}{2}\sum_{ajk}\frac{\left|\bra{Ia}\ket{jk}\right|^2}{-\text{IP}_I + \varepsilon_a-\varepsilon_j-\varepsilon_k}\\ -& \frac{1}{2} \sum_{abj}\frac{\left| \bra{Ij}\ket{ab}\right|^2}{-\text{IP}_I + \varepsilon_j -\varepsilon_a -\varepsilon_b}
\end{aligned}
\end{equation}

\noindent where indices $a,b$ refer to virtual and $j,k$ to occupied HF spin-orbitals. Note that the antisymmetrized 4-center integrals $\bra{Ia}\ket{jk}$ and $\bra{Ij}\ket{ab}$ systematically involve the spin-orbital for which the IP is computed. The above equation can be formally adapted to a DFT reference by constructing the generalized Fock matrix based on the KS orbitals. The occupied and virtual orbitals are then separately rotated such that they become canonical with respect to the generalized Fock matrix. These pseudocanonical orbitals and corresponding new eigenvalues are then used in equation (\ref{soep}). This is known as the GW2X method\cite{shigeta2001electron}. Alternatively, we also propose the GW2X* method, in which KS orbitals are directly used:

\begin{equation}\label{gw2x*}
\begin{aligned}
     \text{IP}_I^{\ \text{GW2X*}} = - f_{II} \ -& \frac{1}{2}\sum_{ajk}\frac{\left|\bra{Ia}\ket{jk}\right|^2}{-\text{IP}_I + f_{aa}-f_{jj}-f_{kk}}\\ -& \frac{1}{2} \sum_{abj}\frac{\left| \bra{Ij}\ket{ab}\right|^2}{-\text{IP}_I + f_{jj} -f_{aa} -f_{bb}}
\end{aligned}
\end{equation}

\noindent where $f_{II}$ is the diagonal element of the generalized Fock matrix corresponding to the $\phi_I$ KS orbital. This approach has an efficiency advantage over the original GW2X as potentially expensive orbital rotations are avoided. Its implementation is also simpler. The second order electron propagator method accounts for relaxation effects upon creation of the core hole as well as electron correlation\cite{pickup1973direct}. Moreover, it is free of self-interaction error since the generalized Fock matrix, which is build with 100\% exact exchange, is used. Finally, this is a fully \textit{ab-initio} scheme that does not rely on any empirical parameter. 

Both the GW2X and GW2X* methods were implemented in CP2K, where they can be employed as a correction scheme for core level LR-TDDFT and/or a way of computing core ionization potentials for X-ray photoelectron spectroscopy (XPS). For each excited core state $\phi_I$ in the system, the RI 3-center ERIs $(pq|\nu)$ and $(\mu|rI)$ are first computed. It is followed by contractions steps from AOs to MOs, such that the anti-symmetrized ERIs needed in equations (\ref{soep}) or (\ref{gw2x*}) can be constructed. The electron propagator equation is then solved using a Newton-Raphson scheme and the resulting IP is substituted in equation (\ref{matrix_A}). From there, the normal LR-TDDFT problem is solved, reusing the precomputed ERIs $(pq|\nu)$ and $(\mu|rI)$. Note that the size of the RI basis $\{\chi_\mu(\mathbf{r})\}$ is independent of system size. Hence, storing $(pq|\mu)$ scales as $\mathcal{O}(n^2)$ in memory at worst. Moreover, contracting $(pq|\mu)$ to \textit{e.g.} $(ab|\mu)$ has the same computational scaling as a normal matrix-matrix multiplication, namely $\mathcal{O}(n^3)$. Since MOs are not localized, storing a fully contracted tensor $(ab|Ij)$ would scale cubically in memory. This can be avoided by contracting $(ab|\mu)$ and $(\nu|Ij)$ first and leaving the RI contraction as the last step. The final contraction can then be done in batches and the $(ab|Ij)$ integrals never fully stored. All integral storage and contraction is done using the sparse matrix and tensor library DBCSR\cite{dbcsr}.

\begin{table*}
\caption{\label{table:l_edge} Comparison of GW2X corrected LR-TDDFT for XANES L-edge spectroscopy with similar methods from the literature. The difference between calculated and experimental first excitation at the L\textsubscript{2}- and L\textsubscript{3}-edge is reported (in eV). All methods include relativistic treatments for the spin--orbit coupling.  The bold and italic atom in the chemical formula is the one from which the excitation takes place.}
\begin{ruledtabular}
\begin{tabular}{l c c c c p{0.02cm} c c c c}
& \multicolumn{4}{c}{\textbf{L\textsubscript{3}}} & & \multicolumn{4}{c}{\textbf{L\textsubscript{2}}} \\ 
\cline{2-5} \cline{7-10} \rule{0pt}{2.6ex} & SCAN-ROKS\cite{hait2020highly} & EOM-CCSD\cite{vidal2020equation} & PBEh-GW2X & Exp. & & SCAN-ROKS\cite{hait2020highly} & EOM-CCSD\cite{vidal2020equation} & PBEh-GW2X & Exp. \\ 
 & \small{aug-cc-pCVTZ} & \small{uC-6-311(2+,+)G** } & \small{aug-pcX-2}& & & \small{aug-cc-pCVTZ} & \small{uC-6-311(2+,+)G** } & \small{aug-pcX-2}& \\
\hline 
\rule{0pt}{2.6ex}\textit{\textbf{Si}}H\textsubscript{4} &  + 0.4 &  - 0.3 &  + 1.2 & 102.6\cite{hayes1972absorption} & &  + 0.4 &  - 0.3 &  + 1.1 & 103.2\cite{hayes1972absorption}\vspace{0.1cm}\\ 
\textit{\textbf{Si}}Cl\textsubscript{4} &  + 0.3 & --- &  - 0.4 & 104.2\cite{bozek1987high} & &  + 0.3 & --- &  - 0.3 & 104.8\cite{bozek1987high}\vspace{0.1cm}\\ 
\textit{\textbf{P}}H\textsubscript{3} &  + 0.2 & --- &  + 0.4 & 131.9\cite{liu1990high} & &  + 0.1 & --- &  + 0.5 & 132.8\cite{liu1990high}\vspace{0.1cm}\\ 
\textit{\textbf{P}}F\textsubscript{3} &  0.0 & --- &  + 0.2 & 134.9\cite{neville1998inner} & &  + 0.1 & --- &  + 0.5 & 135.6\cite{neville1998inner}\vspace{0.1cm}\\ 
H\textsubscript{2}\textit{\textbf{S}} &  + 0.3 &  - 0.2 &  + 0.3 & 164.4\cite{guillemin2005fragmentation} & &  + 0.3 &  - 0.1 &  + 0.2 & 165.6\cite{guillemin2005fragmentation}\vspace{0.1cm}\\ 
OC\textit{\textbf{S}} &  + 0.1 &  + 0.1 &  + 0.1 & 164.3\cite{ankerhold1997ionization} & &  + 0.2 &  - 0.1 &  + 0.2 & 165.5\cite{ankerhold1997ionization}\vspace{0.1cm}\\ 
\textit{\textbf{S}}O\textsubscript{2} & --- &  + 0.1 &  0.0 & 164.4\cite{krasnoperova1976fine} & & --- &  + 0.3 &  + 0.2 & 165.6\cite{krasnoperova1976fine}\vspace{0.1cm}\\ 
H\textit{\textbf{Cl}} &  + 0.1 & --- &  + 0.4 & 200.9\cite{aksela1990decay} & &  + 0.2 & --- &  + 0.6 & 202.4\cite{aksela1990decay}\vspace{0.1cm}\\ 
\textit{\textbf{Cl}}\textsubscript{2} &  0.0 & --- &  - 0.6 & 198.2\cite{nayandin2001angle} & &  + 0.1 & --- &  - 0.5 & 199.8\cite{nayandin2001angle}\vspace{0.1cm}\\ 
\hline 
\rule{0pt}{2.6ex}\textbf{MAD} & 0.2& 0.2& 0.4 & && 0.2& 0.2& 0.4& \\ 
\end{tabular}
\end{ruledtabular}

\end{table*}

\section{Benchmarks \& Results}
The implementation of the GW2X method for core states is tested on a wide range of systems, functionals and basis sets. Benchmarks cover GW2X as a correction to LR-TDDFT for K- and L-edge spectroscopy and as a mean of calculating ionization potentials. We investigate basis set convergence and compare the method to literature benchmarks for molecules. We also discuss the impact of various approximations to accuracy and apply the method to extended systems in periodic boundary conditions.

\subsection{GW2X as correction to LR-TDDFT } 

\begin{figure}[b]
    \centering
    \includegraphics[width=0.48\textwidth]{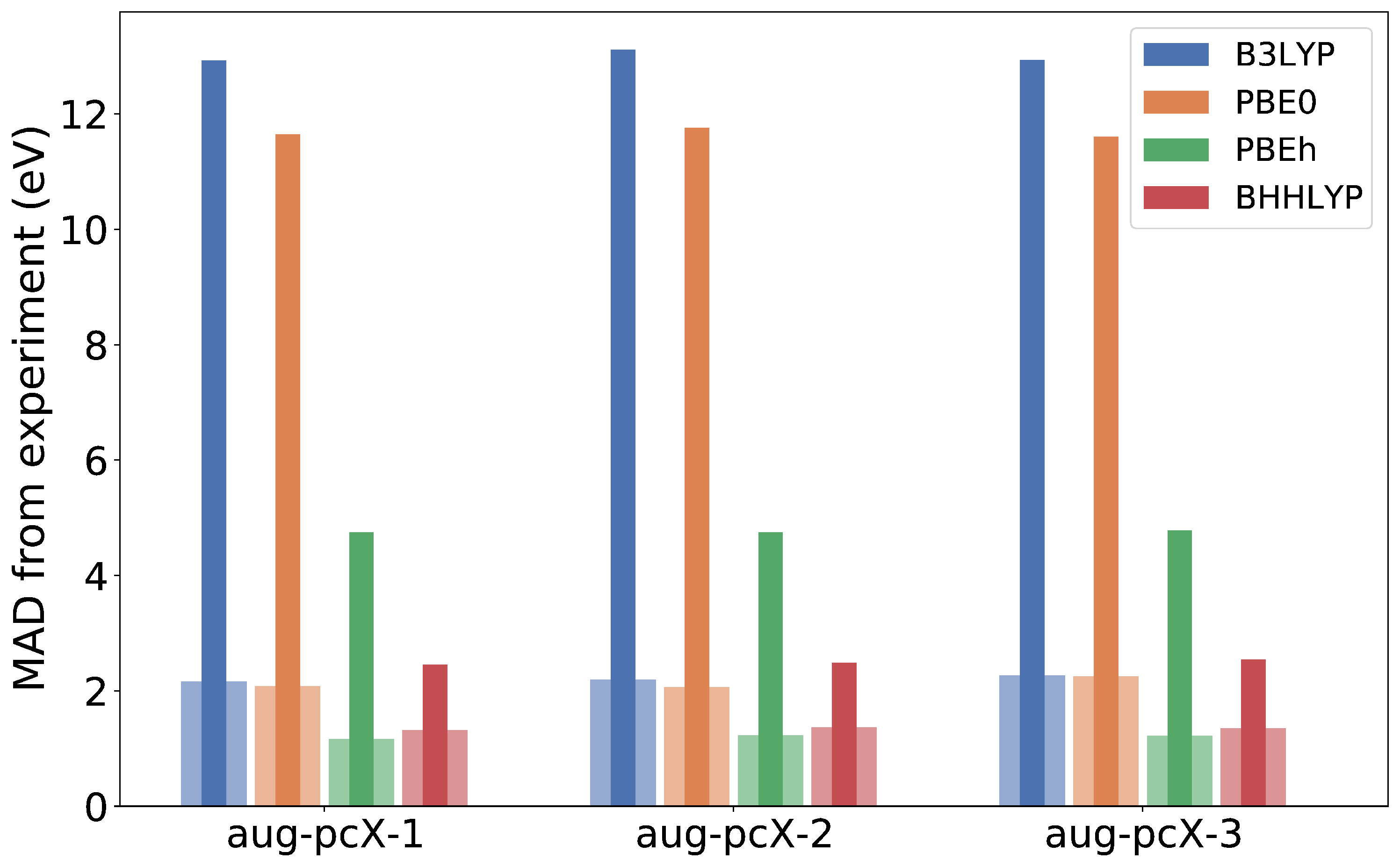}
    \caption{Mean absolute deviation of the first K-edge excitation energy calculated using pure LR-TDDFT (thin dark bars) and GW2X corrected LR-TDDFT (wide clear bars) with respect to experimental values. The benchmark set is made of 25 first and second row atom molecules and covers 43 distinct excitations.}
    \label{fig:k_edge}
\end{figure}

To asses the applicability of the GW2X method as a correction to core LR-TDDFT, we calculated the first K-edge excitation energy of 25 molecules with first and second row atoms and compared to experimental data. The benchmark set includes 16 distinct excitations from molecular C1s levels, 10 from N1s, 13 from O1s and 4 from F1s. The structures were optimized at the def2-TZVP\cite{weigend2005balanced}/B3LYP\cite{stephens1994ab} level. For this benchmark, we used the core-specific aug-pcX basis set\cite{ambroise2018probing} and four common hybrid functionals with increasing fraction of Hartree--Fock exchange; B3LYP(20\%), PBE0(25\%)\cite{adamo1999toward} PBEh(45\%)\cite{atalla2013hybrid} and BHHLYP(50\%)\cite{becke1993new}. On figure \ref{fig:k_edge}, the mean absolute deviation (MAD) of LR-TDDFT and LR-TDDFT+GW2X first excitation energies with respect to experiments is displayed for the different functional and basis set combinations. The GW2X correction systematically improves the LR-TDDFT results and brings down the very disparate MADs of the different functionals to a similar level. This can be explained by the fact that a higher HFX fraction leads to a reduction of the self-interaction error whereas the GW2X correction is free of it altogether. It can also be observed that increasing the basis set quality from double to quadruple zeta quality does not significantly change the results, suggesting rapid convergence. Finally, the GW2X corrected PBEh($\alpha=0.45$) functional performs the best overall with MADs of 1.32, 1.38 and 1.37 eV for the aug-pcX-1, aug-pcX-2 and aug-pcX-3 basis sets, respectively. A detailed table containing the calculated and reference energies for each molecule is available in the supplementary material. 

\begin{table*}
\caption{\label{table:ip} Comparison of GW2X 1s core ionization potentials with similar methods from the literature. The difference between the calculated and experimental IP is reported (in eV). The bold and italic atom in the chemical formula is the one being ionized.}

\begin{ruledtabular}
\begin{tabular}{l c c c c c c c c c}
 & B3LYP-$\Delta$SCF\cite{besley2009self} & PBEh-G0W0\cite{golze2020accurate} & EOM-CCSD\cite{vidal2019new} & EP2\cite{flores2007assessment} & TOEP2\cite{flores2007assessment} & PBEh-GW2X & Exp. \\
 & \small{u6-311G**} & \small{cc-pVQZ} & \small{aug-cc-pVTZ} & \small{cc-pVTZ} & \small{cc-pVTZ} & \small{aug-pcX-2}& \\
\hline
\rule{0pt}{2.6ex}\textit{\textbf{C}}H\textsubscript{4} &  + 0.17 &  - 0.44 & --- &  + 0.75 &  + 0.31 &  + 0.09 & 290.84\cite{myrseth2002adiabatic}\vspace{0.15cm}\\
\textit{\textbf{C}}\textsubscript{2}H\textsubscript{4} & --- &  - 0.28 &  + 0.39 &  + 0.99 &  + 1.37 &  + 0.18 & 290.82\cite{myrseth2002adiabatic}\vspace{0.15cm}\\
\textit{\textbf{C}}O &  + 0.51 &  - 0.65 &  + 0.45 &  + 1.74 &  + 1.10 &  + 0.91 & 296.23\cite{myrseth2002adiabatic}\vspace{0.15cm}\\
\textit{\textbf{C}}H\textsubscript{2}O &  + 0.31 &  - 0.11 & --- &  + 1.59 &  + 0.93 &  + 0.76 & 294.38\cite{bakke1980table}\vspace{0.15cm}\\
\textit{\textbf{N}}H\textsubscript{3} &  - 0.01 &  - 0.61 &  + 0.60 & --- & --- &  - 1.04 & 405.52\cite{bakke1980table}\vspace{0.15cm}\\
\textit{\textbf{N}}NO &  + 0.21 & --- & --- &  + 0.29 &  + 0.44 &  - 1.04 & 408.66\cite{jolly1984core}\vspace{0.15cm}\\
N\textit{\textbf{N}}O &  + 0.24 & --- & --- &  + 0.75 &  + 1.76 &  - 0.57 & 412.57\cite{jolly1984core}\vspace{0.15cm}\\
C\textit{\textbf{O}} &  + 0.33 &  - 0.92 &  + 1.33 &  - 0.93 &  - 0.24 &  - 2.17 & 542.10\cite{siegbahn1969applied}\vspace{0.15cm}\\
H\textsubscript{2}\textit{\textbf{O}} &  - 0.41 &  - 1.30 &  + 0.75 &  - 2.02 &  - 0.42 &  - 2.48 & 539.90\cite{jolly1984core}\vspace{0.15cm}\\
CH\textsubscript{2}\textit{\textbf{O}} &  - 0.19 &  - 1.12 & --- &  - 1.93 &  - 0.83 &  - 2.27 & 539.33\cite{bakke1980table}\vspace{0.15cm}\\
H\textit{\textbf{F}} &  - 0.45 & --- & --- &  - 3.44 &  - 0.98 &  - 3.55 & 694.18\cite{jolly1984core}\vspace{0.15cm}\\
\textit{\textbf{F}}\textsubscript{2} &  - 0.38 & --- & --- &  - 2.42 &  - 1.28 &  - 3.02 & 696.69\cite{jolly1984core}\vspace{0.15cm}\\
\hline
\rule{0pt}{2.6ex}\textbf{MAD} & 0.29& 0.68& 0.70& 1.53& 0.88& 1.51& \\
\end{tabular}
\end{ruledtabular}

\end{table*}

In table \ref{table:k_edge}, LR-TDDFT with GW2X correction is compared to other XANES K-edge calculation methods reported in the literature. A smaller selection of nine molecules (12 excitations) was made such as to maximize overlap with the other studies. The aug-pcX-2 basis (triple zeta quality) was chosen for similar reasons and the  PBEh($\alpha$=0.45) functional selected because it performed best in the previous benchmark. Note that the reported MADs can only be used as semi-quantitative measure of method quality because of the small sample size, patchy data and varying basis sets. GW2X corrected LR-TDDFT performs on par with B3LYP-$\Delta$SCF and equation of motion CCSD, while surpassing CIS(D). SRC2-TDDFT performs better, but the core-specific range separated SRC2 functional has four parameters which were specifically fitted on those molecules\cite{besley2009time}. Finally, the square gradient minimization method (SGM) developed by Hait and Head-Gordon\cite{hait2020excited} combined with ROKS and the SCAN\cite{sun2015strongly} functional also yields better results. However, both LR-TDDFT and the GW2X correction scheme can be applied without prior knowledge of a target system whereas SGM requires careful preparation of an initial guess. This "black-box" aspect of LR-TDDFT+GW2X makes it an interesting method for high-throughput studies.

GW2X can be used to correct XANES LR-TDDFT at the L-edge, where the ionization potential of the three 2p states are computed and spin--orbit coupling is included at the TDDFT level. It is however necessary to compare calculated and experimental spectra to reliably extract the first excitation energies at the L\textsubscript{2} and L\textsubscript{3} peaks, making large scale benchmark studies difficult. Instead, we focus on a few molecules present in similar studies and compare results in table \ref{table:l_edge}. Geometries were optimized at the def2-TZVP/B3LYP level and the PBEh($\alpha=0.45$) functional with aug-pcX-2 basis chosen for the LR-TDDFT+GW2X calculations. All three methods perform similarly well, both at the L\textsubscript{2} and L\textsubscript{3} peaks, and with errors comparable to those observed at the K-edge. Note that on average, the GW2X correction induces a blue shift of 2.4 eV compared to bare LR-TDDFT, again systematically improving the results. The amount of spin--orbit splitting is well captured by all three methods.

\subsection{GW2X for core IP calculations}

\begin{figure}[b]
    \centering
    \includegraphics[width=0.48\textwidth]{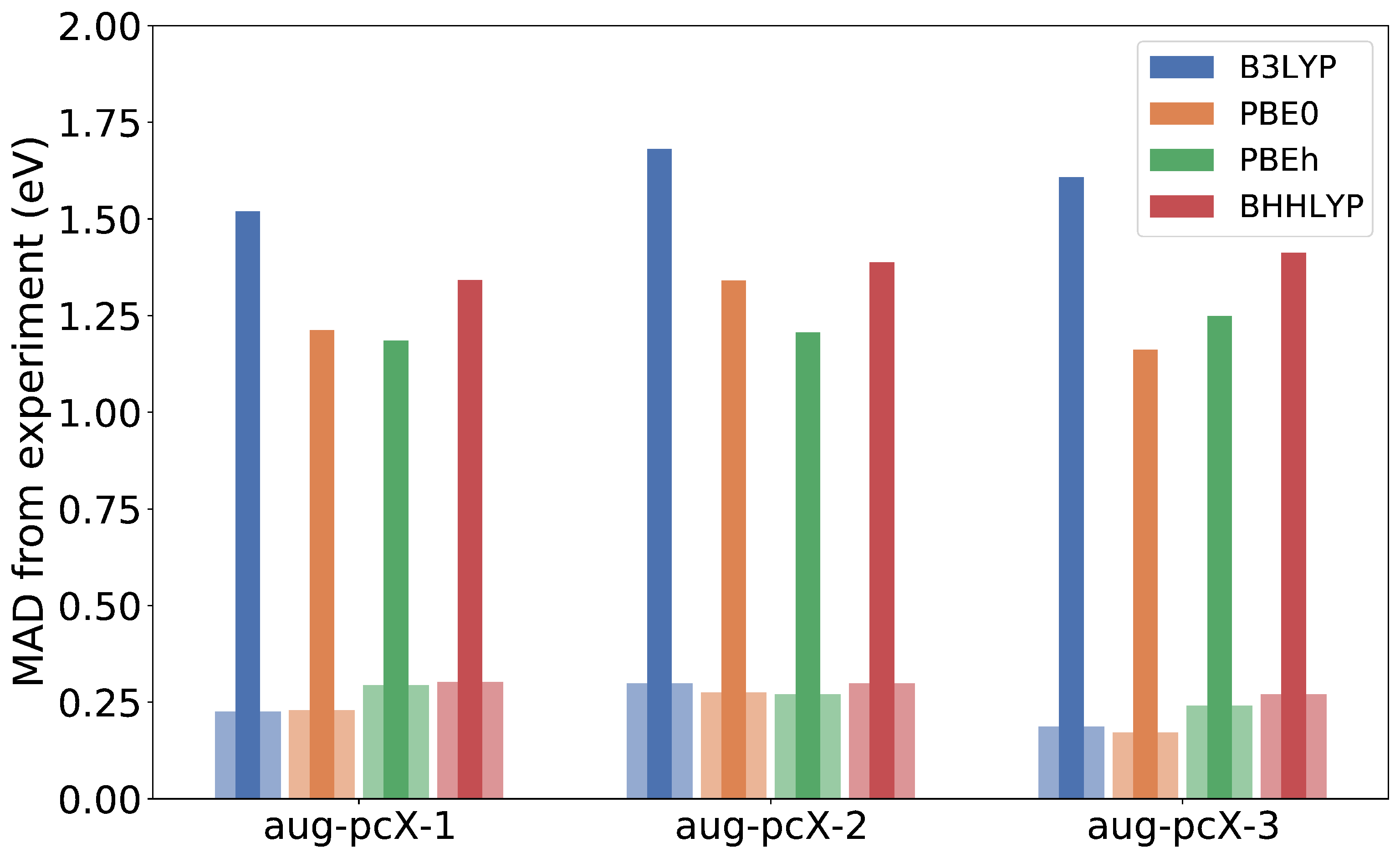}
    \caption{Mean absolute deviation of the core 1s ionization potentials calculated with GW2X over the CORE65 benchmark set with respect to experiment. Thin dark bars represent the absolute IPs and wide clear bars the relative IPs.}
    \label{fig:ip_conv}
\end{figure}

GW2X can also be used as a stand alone method to compute ionization potentials. In this case, all TDDFT related operations can be ignored for efficiency. Alternatively, the IPs can be obtained as a side-product of a corrected LR-TDDFT calculation. To asses the quality of GW2X in this context, the CORE65 benchmark set\cite{golze2020accurate} was used. It contains 65 distinct core states over 32 molecules with first and second row atoms, covering 30 C1s, 21 O1s, 11 N1s and 3 F1s. Core ionization potentials were calculated with the B3LYP, PBE0, PBEh($\alpha=0.45$) and BHHLYP functionals and the aug-pcX basis sets. Mean average deviations from experiments are displayed in figure \ref{fig:ip_conv}, both for absolute and relative energies. Note that relative IPs are defined as the difference with respect to a reference, which was taken to be methane for C1s, ammonia for N1s, water for O1s and methyl fluoride for F1s. As observed for K-edge excitation energies, basis set quality does not significantly affect the MAD from experiments, suggesting fast basis set convergence. All functionals perform similarly, with a slight edge for the PBE flavored hybrids. The error on the absolute IPs stands at 1.35 $\pm$ 0.16 eV, which is very close to the errors observed at the K-edge. Relative IPs are remarkably well reproduced, which means that core binding energies from different systems can be reliably compared. Detailed values can be found in the supplementary material.

In table \ref{table:ip}, core 1s ionization potentials calculated using GW2X and similar methods from the literature are displayed. The choice of molecules is such that the overlap with other works is maximized. As discussed earlier, missing data and different basis sets only allow for a semi-quantitative comparison. It is noteworthy that B3LYP-$\Delta$SCF performs the best, despite being the conceptually simplest method. The higher order methods equation-of-motion CCSD, PHEh($\alpha=0.45$)-G0W0 and transition operator electron propagator (TOEP2) yield comparable results with each other and are relatively close to B3LYP-$\Delta$SCF. The second-order electron propagator method (EP2) and GW2X perform remarkably similarly. This is to be expected since GW2X is the DFT version of EP2, but this nonetheless underlines the validity of the generalization scheme. Despite exhibiting lower accuracy than the other methods, our implementation of GW2X still improves upon Koopmans' theorem\cite{koopmans1934zuordnung} by an order of magnitude\cite{flores2007assessment} while only scaling cubically. Note that the GW2X core IPs increasingly diverge from experiments with atomic number. This behavior is however not observed for the GW2X LR-TDDFT corrected K-edge energies of table \ref{table:k_edge}, which suggests that some  form of error cancellation takes place.

\subsection{Approximations for increased efficiency}\label{approx}

Various approximations can be employed to speed up GW2X calculations. The auxiliary density matrix method\cite{guidon2010auxiliary} under its purified (ADMM1) and non-purified flavors (ADMM2) allow for very efficient ground state hybrid functional calculations. We previously showed in ref. ~\onlinecite{bussy2021efficient} that such an approximate calculation can serve as a base to a LR-TDDFT perturbative treatment with only minor loss of accuracy. In this work, we further propose the ADMM method as a mean to accelerate the construction of the generalized Fock matrix, which is a necessary step of the GW2X method. In table \ref{table:admm}, we investigate the impact of the ADMM approximation on K-edge corrections and core IP calculations compared to the full Hartree--Fock exchange treatment. Additionally, we report the results obtained using the GW2X* method (as described in section \ref{theory}) and the non core-specific aug-pcseg\cite{jensen2014unifying}/aug-admm\cite{kumar2018accelerating} basis sets. There is practically no difference of accuracy between the two ADMM schemes, making the more efficient non-purified version the default choice. Compared to full HFX, ADMM introduces an additional error of 0.2-0.3 eV for GW2X K-edge correction and 0.4-0.5 eV for core IPs. The error is increased when ADMM is used together with the more approximate GW2X* scheme, especially for IP calculations. It is noteworthy that the use of the general purpose aug-pcseg-2 and corresponding aug-admm-2 basis sets do not greatly change the results, suggesting that large core-specific basis sets are not strictly necessary in this context. This is particularly useful for making large scale calculations more affordable. Finally, the errors induced by the  ADMM and/or GW2X* approximations at the K-edge are rather small compared to the scale of a near-edge X-ray absorption spectrum, which typically spans 15-20 eV. 

\begin{table}
\caption{\label{table:admm} Mean average deviation (in eV) of the GW2X method with respect to experiments, using various approximations. The same benchmark sets as for figures \ref{fig:k_edge} and \ref{fig:ip_conv} were used for K-edge and core ionization potentials, respectively. The PBEh($\alpha=0.45$) functional as well as the aug-pcX-2/aug-admm-2 and aug-pcseg-2/aug-admm-2 basis set combinations were used. The results within parenthesis refer to the latter.                }
\begin{ruledtabular}
\begin{tabular}{l c c p{0.01cm} c c}
& \multicolumn{2}{c}{K-edge} & & \multicolumn{2}{c}{IP}  \\ 
\cline{2-3} \cline{5-6} \rule{0pt}{2.6ex}
& GW2X & GW2X* & & GW2X & GW2X* \\ 
\hline 
 \rule{0pt}{2.6ex}Full HFX & 1.2 (1.2) & 1.3 (1.2)&& 1.2 (1.0) & 1.6 (1.6) \vspace{0.1cm}\\ 
ADMM1 & 1.4 (1.3) & 1.6 (1.4) && 1.6 (1.4) & 2.2 (1.8) \vspace{0.1cm}\\ 
ADMM2 & 1.5 (1.3) & 1.6 (1.5) && 1.7 (1.4) & 2.2 (1.9) \\ 
\end{tabular}
\end{ruledtabular}

\end{table}

\subsection{GW2X in extended systems}

\begin{figure*}[tb]
    \centering
    \includegraphics[width=1.0\textwidth]{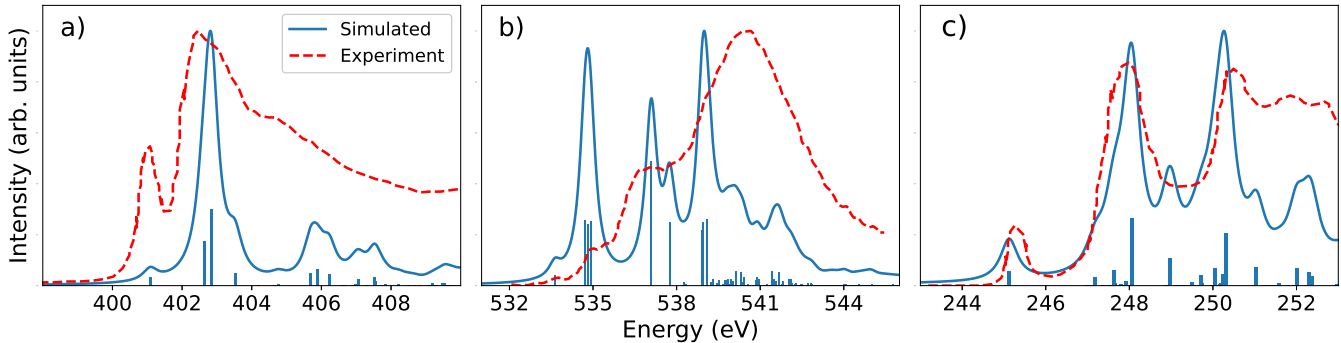}
    \caption{GW2X corrected LR-TDDFT spectra of a) solid ammonia N K-edge, b) ice 1h O K-edge and c) solid argon Ar L\textsubscript{2,3}-edge in periodic boundary conditions. The PBEh($\alpha=0.45$) functional and a mix of all electron basis sets and pseudopotentials were used. The GW2X corrections amount to blue shifts of 3.7 eV, 2.8 eV and 3.5 eV for ammonia, ice and argon, respectively. A Lorentzian broadening of fwhm 0.5 eV was applied and calculated intensities uniformly scaled to match experiments. Experimental data come from refs ~\onlinecite{parent2009irradiation}, ~\onlinecite{wernet2004structure} and ~\onlinecite{haensel1971optical}.}
    \label{fig:solids}
\end{figure*}

There is, in principle, no issue with using GW2X as a correction scheme in periodic boundary conditions (PBCs). Moreover, the high levels of symmetry found in crystal structures can be exploited for efficiency. Since multiple atoms are equivalent, it is only necessary to correct the core KS eigenvalue for one of those, leading to a $\mathcal{O}(n^3)$ scaling with system size. Note that disordered systems such as liquids are limited to a $\mathcal{O}(n^4)$ scaling, as each atom may have a slightly different environment. GW2X being related to second-order M\o ller-Plesset perturbation theory\cite{moller1934note}, it suffers from the same limitations. In particular, it works best for large gap molecular systems whereas semiconductors require large supercells for converged results\cite{gruber2018applying}.

We applied the GW2X correction schemes on three molecular crystals, namely solid ammonia, ice 1h and solid argon. The respective unit cells contain 128, 288 and 32 atoms and the structures were first relaxed at the DZVP-MOLOPT-SR-GTH\cite{vandevondele2007gaussian}/PBE\cite{perdew1996generalized}+D3\cite{grimme2010consistent} level of theory. LR-TDDFT+GW2X calculations were then performed using the PBEh($\alpha=0.45$) hybrid functional with the truncated Coulomb potential\cite{guidon2009robust} (truncation radii of 5 \AA, 6 \AA\ and 5\AA, \textit{i.e.} just under half the cell parameter). The corrected LR-TDDFT spectra are shown in figure \ref{fig:solids}. The first two absorption peaks of crystalline NH3 are well aligned to the experimental spectrum, thanks to the GW2X correction that amounts to a blue shift of 3.7 eV. The relative intensity of the first peak is too small, but this issue lies with LR-TDDFT rather than the correction scheme. The triple zeta aug-pceseg-2/aug-admm-2 basis sets were used to describe a single excited nitrogen atom while all other atoms relied on GTH pseudopotentials\cite{goedecker1996separable, hartwigsen1998relativistic, krack2005pseudopotentials} and DZVP-MOLOPT-SR-GTH/FIT3\cite{guidon2010auxiliary} basis sets. The ice 1h spectrum was calculated using the same basis sets, where only one excited oxygen atom was described at the all electron level as well. The GW2X correction to the spectrum is a blue shift of 2.8 eV, which also aligns the first few peaks remarkably well. Similarly to the NH3 calculation, the relative intensities of the calculated spectrum do not perfectly match experimental data. The solid argon simulated spectrum is well aligned (GW2X shift of 3.5 eV) and enjoys good relative intensities. Using the higher quality quadruple zeta aug-pcseg-3 was found to be necessary and the added diffuse functions of the augmented set especially crucial. Only one atom was described at the all electron level and all others using DZVP-MOLOPT-SR-GTH basis sets. However, since there is no available auxiliary FIT3 basis set for argon, full HFX was employed instead of the ADMM approximation. Note that using quadruple zeta basis sets for ammonia and ice did not improve the results. 

\begin{figure}[b]
    \centering
    \includegraphics[width=0.475\textwidth]{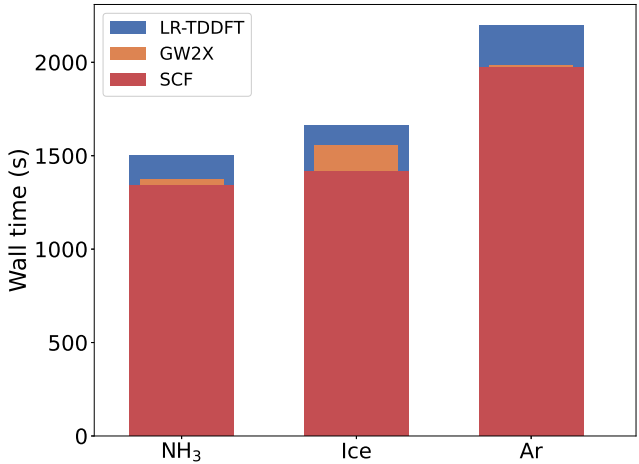}
    \caption{Execution times of the extended systems calculations of figure \ref{fig:solids}. All calculations were carried out on an Intel S2600WT2R machine with 24 CPUs and 256 GB of memory. Different colors represent time spent treating the ground state SCF, the LR-TDDFT equations and the GW2X correction, respectively. Note that the time spent evaluating the AOs RI integrals shared by GW2X and LR-TDDFT are credited to the latter. }
    \label{fig:timings}
\end{figure}

Most of the efficiency aimed approximations discussed in section \ref{approx} were used in our extended system calculations. General purpose aug-pcseg basis sets were used throughout as was the non-purified ADMM scheme for solid NH3 and ice 1h. Moreover, only one atom per system was described at the all electron level while all others relied on pseudopotentials, essentially freezing their core and reducing the number of MOs to include in equation (\ref{soep}). Only the original GW2X scheme was preferred over the simplified GW2X* version as the systems are too small for notable efficiency gains. The spectra obtained with the latter method are available in the supplementary material. Note that a truncated Coulomb potential was used for exact exchange integrals in order to avoid nonphysical self-exchange\cite{guidon2009robust}. The same operator was used for ground state HFX, LR-TDDFT exact exchange kernel, and thus also GW2X. The use of such a short range exchange operator also reduces the cost and scaling of the previously mentioned integrals. We did not observe a dependence of the GW2X results on truncation radius, provided that it is large enough (5-6 \AA). To illustrate the overall affordability of the method, all calculations were run on a 24 CPU cores system and execution times reported on figure \ref{fig:timings}. The ground state SCF calculation dominates overall, mostly because of the ERIs evaluation for the HFX fraction of the functional. This is so even with the initial SCF guess coming from a converged PBE calculation and the ADMM approximation for ammonia and ice. The efficiency of the GW2X correction is emphasized by the fact that it takes at most ~ 60\% of the post-SCF effort in ice, and is almost negligible for the other two smaller systems. The most expensive calculation, solid argon, ran with a wall time of just under 40 minutes. 

In periodic systems, GW2X cannot be directly used to calculate absolute binding energies. Because the potential reference can be arbitrarily defined, whereas it goes to zero far away in the non-periodic case, KS eigenvalues may be uniformly shifted. Thus, the solution of equation (\ref{soep}) for the IP will be shifted by some unknown amount in PBCs. However, once plugged into equation (\ref{tddft_simple}) for the LR-TDDFT correction, the shifts cancel out. A possible way to calculate core IPs in PBCs using GW2X would be to use slab models and reference the KS eigenvalues against the vacuum level, as it is done in the \textit{GW} community\cite{chen2014band, pham2014probing, hinuma2014band}.

\section{Conclusion}
         
We have implemented a correction scheme to address the lack of orbital relaxation and the self-interaction error that afflicts the prediction of absolute excitation energies in LR-TDDFT for core level spectroscopy. The methods is based on the DFT generalization of the second-order electron propagator theory (GW2X). It allows for the accurate calculation of ionization potentials, which are then substituted in the LR-TDDFT equations, replacing the KS eigenvalues. The implementation exploits the local nature of core states for efficiency, scaling cubically with system size and integrating seamlessly in our existing  core level LR-TDDFT implementation.

Benchmarks at the K- and L-edges show that the GW2X correction scheme systematically improves LR-TDDFT results for four common hybrid functionals, namely B3LYP, PBE0, PBEh($\alpha$=0.45) and BHHLYP, reaching accuracies comparable to higher level methods such as equation of motion CCSD. Moreover, the method can be used to calculate core ionization potentials directly with similar errors to those observed at the K-edge. Rapid basis set convergence was observed in both LR-TDDFT correction and core binding energy calculation and the use of core-specific basis sets does not seem essential.

The method is applicable with periodic boundary conditions for large gap  molecular  systems, as demonstrated for solid ammonia, ice 1h and crystalline argon. Such calculations are accessible with modest computer resources and are made feasible due to efficient approximations. In particular, mixing the ADMM scheme and a hybrid all-electron/pseupopotential description has proven effective. Core ionization potentials are however not easily accessible in such systems and would require additional calculations.

This work focuses on standard GW2X and there is room for further exploration. The initial GW2X paper\cite{shigeta2001electron} also describes the GW(2) approach, which could prove more efficient since the exchange contributions are ignored. Alternatively, accuracy could  potentially be improved by using spin scaling approaches, as it is done in SCS- and SOS-MP2\cite{grimme2005accurate,jung2004scaled}.

\section{Acknowledgement}

This work is supported by the MARVEL National Centre for Competency in Research funded by the Swiss National Science Foundation (grant agreement ID 51NF40-182892).

\nocite{*}
\bibliography{main}

\end{document}


\title{Supplementary material for:  First principles correction scheme for linear-response time-dependent density functional theory calculations of core electronic states}

\author{Augustin Bussy}%
 \email{augustin.bussy@chem.uzh.ch}
\altaffiliation{Department of Chemistry, University of Zurich, Winterthurerstrasse 190, CH-8057 Zürich, Switzerland.}%
\author{J\"urg Hutter}%
 \email{hutter@chem.uzh.ch}
\affiliation{Department of Chemistry, University of Zurich, Winterthurerstrasse 190, CH-8057 Zürich, Switzerland.
}%

\date{\today}
\maketitle


\section{K-edge benchmark tables}

The detailed values of the benchmark studies of GW2X corrected LR-TDDFT for core excitations are given in a table.  Due to the large amount of entries, the table was split in halves: table \ref{K1} and \ref{K2}. As a reminder, geometries were optimized at the B3LYP/def2-TZVP level. All results reported in these tables were obtained using the triple zeta aug-pcX-2 basis set.

\section{Core IP benchmark tables}

The core IP values obtained using the GW2X method are given in tables \ref{IP1}, \ref{IP2} and \ref{IP3}. The molecule structures come straight from the CORE65 benchmark set\cite{golze2020accurate}. The triple zeta aug-pcX-2 basis set is used throughout.

\begin{table*}  
\caption{Part I of the K-edge correction benchmark table. }\label{K1}
\begin{ruledtabular}
\begin{tabular}{l c c c c p{0.02cm} c c c c p{0.02cm} c}
& \multicolumn{4}{c}{LR-TDDFT}  && \multicolumn{4}{c}{LR-TDDFT+GW2X} &&  \\ 
\cline{2-5} \cline{7-10} \rule{0pt}{2.6ex}
& B3LYP & PBE0  & PBEh  & BHHLYP  & & B3LYP & PBE0  & PBEh & BHHLYP  && Exp. \\ 
\hline 
     \rule{0pt}{2.6ex}\textit{\textbf{C}}H\textsubscript{4}& -11.7& -10.5& -4.0& -1.9 & & +2.5& +2.6& +2.0& +1.6&&  288.0\cite{schirmer1993k} \vspace{0.1cm}\\ 
    \textit{\textbf{C}}\textsubscript{2}H\textsubscript{2}& -10.7& -9.7& -4.2& -2.3 & & +3.0& +3.0& +1.3& +0.8&&  285.9\cite{hitchcock1977carbon} \vspace{0.1cm}\\ 
    \textit{\textbf{C}}\textsubscript{2}H\textsubscript{4}& -10.2& -9.5& -4.2& -2.2 & & +3.4& +3.0& +1.2& +0.8&&  284.7\cite{hitchcock1977carbon} \vspace{0.1cm}\\ 
    \textit{\textbf{C}}\textsubscript{2}H\textsubscript{6}& -10.6& -10.2& -3.7& -1.3 & & +3.1& +2.5& +1.9& +1.8&&  286.9\cite{hitchcock1977carbon} \vspace{0.1cm}\\ 
    H\textit{\textbf{C}}HO& -10.3& -9.7& -4.6& -2.6 & & +4.5& +4.2& +2.1& +1.6&&  285.6\cite{remmers1992high} \vspace{0.1cm}\\ 
    H\textit{\textbf{C}}N& -10.7& -9.8& -4.5& -2.5 & & +3.9& +3.7& +1.9& +1.4&&  286.4\cite{hitchcock1979inner} \vspace{0.1cm}\\ 
    \textit{\textbf{C}}\textsubscript{2}N\textsubscript{2}& -11.0& -10.3& -4.7& -2.5 & & +2.8& +2.5& +0.9& +0.7&&  286.3\cite{hitchcock1979inner} \vspace{0.1cm}\\ 
    \textbf{\textit{C}O}& -11.3& -10.5& -5.7& -3.8 & & +5.1& +4.8& +2.5& +1.9&&  287.4\cite{domke1990carbon} \vspace{0.1cm}\\ 
    \textit{\textbf{C}}O\textsubscript{2}& -11.0& -10.3& -5.3& -3.4 & & +4.6& +4.2& +2.1& +1.6&&  290.8\cite{prince1999vibrational} \vspace{0.1cm}\\ 
    \textit{\textbf{C}}H\textsubscript{3}OH& -11.2& -10.2& -3.8& -1.8 & & +3.0& +3.1& +2.3& +1.8&&  288.0\cite{prince2003near} \vspace{0.1cm}\\ 
    H\textit{\textbf{C}}OOH& -10.7& -10.0& -5.1& -3.1 & & +4.2& +3.8& +1.7& +1.3&&  288.1\cite{prince2003near} \vspace{0.1cm}\\ 
    H\textit{\textbf{C}}OF& -10.6& -9.9& -4.9& -2.9 & & +4.6& +4.2& +2.2& +1.7&&  288.2\cite{robin1988fluorination} \vspace{0.1cm}\\ 
    \textit{\textbf{C}}F\textsubscript{2}O& -10.6& -10.0& -5.1& -3.1 & & +5.6& +4.2& +2.0& +1.6&&  290.9\cite{robin1988fluorination} \vspace{0.1cm}\\ 
    H\textsubscript{2}\textit{\textbf{C}}O& -10.8& -10.1& -5.0& -3.1 & & +4.1& +3.8& +1.7& +1.2&&  286.0\cite{remmers1992high} \vspace{0.1cm}\\ 
    Acetone(CO)& -10.4& -9.8& -4.9& -3.0 & & +3.4& +3.0& +0.8& +0.3&&  286.4\cite{prince2003near} \vspace{0.1cm}\\ 
    Acetone(CH\textsubscript{3})& -13.7& -12.3& -4.9& -2.4 & & +0.1& +0.4& +0.7& +0.7&&  287.4\cite{prince2003near} \vspace{0.1cm}\\ 
    \textit{\textbf{N}}\textsubscript{2}& -12.4& -11.2& -5.0& -2.9 & & +3.4& +3.2& +1.0& +0.2&&  400.9\cite{myhre2018theoretical} \vspace{0.1cm}\\ 
    \textit{\textbf{N}}NO& -12.1& -10.9& -4.4& -2.3 & & +2.1& +2.1& +0.2& -0.6&&  401.0\cite{prince1999vibrational} \vspace{0.1cm}\\ 
    N\textit{\textbf{N}}O& -12.2& -11.1& -4.9& -2.7 & & +2.8& +2.6& +0.4& -0.4&&  404.6\cite{prince1999vibrational} \vspace{0.1cm}\\ 
    HC\textit{\textbf{N}}& -12.2& -10.9& -4.5& -2.4 & & +2.3& +2.3& +0.2& -0.6&&  399.7\cite{hitchcock1979inner} \vspace{0.1cm}\\ 
    C\textsubscript{2}\textit{\textbf{N}}\textsubscript{2}& -12.1& -10.8& -4.2& -2.0 & & +1.7& +1.8& +0.0& -0.7&&  398.9\cite{hitchcock1979inner} \vspace{0.1cm}\\ 
\end{tabular}
\end{ruledtabular}
\end{table*}

\begin{table*} 
\caption{Part II of the K-edge correction benchmark table. }\label{K2}
\begin{ruledtabular}
\begin{tabular}{l c c c c p{0.02cm} c c c c p{0.02cm} c}
& \multicolumn{4}{c}{LR-TDDFT}  && \multicolumn{4}{c}{LR-TDDFT+GW2X} &&  \\ 
\cline{2-5} \cline{7-10} \rule{0pt}{2.6ex}
& B3LYP & PBE0  & PBEh  & BHHLYP  & & B3LYP & PBE0  & PBEh & BHHLYP  && Exp. \\ 
\hline 
     Glycine(N)& -13.5& -11.8& -3.7& -1.5 & & +0.9& +1.4& +1.1& +0.3&&  401.2\cite{plekan2007x} \vspace{0.1cm}\\ 
    Pyrrole(N)& -16.5& -14.4& -5.3& -2.7 & & -3.0& -2.3& -1.5& -1.8&&  402.3\cite{pavlychev1995nitrogen} \vspace{0.1cm}\\ 
    Imidazole(CH=\textit{\textbf{N}}-CH)& -14.4& -12.5& -4.6& -2.3 & & -0.9& -0.2& -0.7& -1.3&&  399.9\cite{apen1993experimental} \vspace{0.1cm}\\ 
    Imidazole(CH-\textit{\textbf{N}}H-CH)& -14.8& -12.8& -4.5& -2.3 & & -1.2& -0.5& -0.6& -1.3&&  402.3\cite{apen1993experimental} \vspace{0.1cm}\\ 
    \textit{\textbf{N}}H\textsubscript{3}& -13.3& -11.7& -4.2& -2.0 & & +1.5& +1.8& +0.9& +0.1&&  400.8\cite{schirmer1993k} \vspace{0.1cm}\\ 
    HCH\textit{\textbf{O}}& -14.1& -12.6& -5.1& -2.6 & & +1.1& +1.1& -1.1& -2.1&&  530.8\cite{remmers1992high} \vspace{0.1cm}\\ 
    C\textbf{\textit{O}}& -14.4& -12.8& -5.0& -2.6 & & +1.3& +1.4& -0.6& -1.5&&  534.2\cite{domke1990carbon} \vspace{0.1cm}\\ 
    CH\textsubscript{3}\textit{\textbf{O}}H& -12.3& -10.4& -1.8& +0.6 & & +3.2& +3.6& +2.4& +1.4&&  531.4\cite{prince2003near} \vspace{0.1cm}\\ 
    HC\textit{\textbf{O}}OH& -14.6& -13.0& -5.3& -2.8 & & +0.3& +0.4& -1.5& -2.3&&  532.2\cite{prince2003near} \vspace{0.1cm}\\ 
    HCO\textit{\textbf{O}}H& -15.7& -13.8& -5.1& -2.3 & & -0.4& -0.1& -1.1& -1.6&&  535.4\cite{prince2003near} \vspace{0.1cm}\\ 
    HC\textit{\textbf{O}}F& -14.5& -12.9& -5.3& -2.8 & & +0.9& +0.9& -1.2& -2.1&&  532.1\cite{robin1988fluorination} \vspace{0.1cm}\\ 
    C\textit{\textbf{O}}\textsubscript{2}& -14.5& -12.8& -4.7& -2.2 & & +0.6& +0.8& -0.8& -1.6&&  535.4\cite{prince1999vibrational} \vspace{0.1cm}\\ 
    N\textsubscript{2}\textit{\textbf{O}}& -14.2& -12.4& -4.1& -1.5 & & +0.2& +0.5& -0.9& -1.6&&  534.6\cite{prince1999vibrational} \vspace{0.1cm}\\ 
    H\textsubscript{2}\textit{\textbf{O}}& -15.3& -13.4& -4.8& -2.4 & & +0.7& +1.1& -0.1& -1.1&&  534.0\cite{schirmer1993k} \vspace{0.1cm}\\ 
    H\textsubscript{2}C\textit{\textbf{O}}& -14.1& -12.6& -5.1& -2.6 & & +1.1& +1.1& -1.1& -2.1&&  530.8\cite{remmers1992high} \vspace{0.1cm}\\ 
    CF\textsubscript{2}\textit{\textbf{O}}& -14.3& -12.7& -5.0& -2.4 & & +1.3& +1.3& -0.7& -1.4&&  532.7\cite{robin1988fluorination} \vspace{0.1cm}\\ 
    Acetone(O)& -14.5& -12.9& -5.3& -2.9 & & +0.3& +0.3& -1.8& -2.6&&  531.4\cite{prince2003near} \vspace{0.1cm}\\ 
    Furan(O)& -16.2& -13.9& -5.0& -2.5 & & -1.8& -1.0& -1.8& -2.6&&  535.2\cite{duflot2003core} \vspace{0.1cm}\\ 
    \textit{\textbf{F}}\textsubscript{2}& -15.4& -13.8& -5.9& -3.3 & & +2.7& +2.5& -0.7& -1.9&&  682.2\cite{hitchcock1981k} \vspace{0.1cm}\\ 
    HCO\textit{\textbf{F}}& -18.2& -16.0& -6.0& -2.8 & & -0.9& -0.6& -1.5& -2.1&&  687.7\cite{robin1988fluorination} \vspace{0.1cm}\\ 
    HF& -18.0& -15.9& -6.4& -3.6 & & +0.4& +0.8& -0.8& -1.8&&  687.4\cite{hitchcock1981k} \vspace{0.1cm}\\ 
    C\textit{\textbf{F}}\textsubscript{2}O& -18.3& -16.1& -6.0& -2.9 & & -0.5& -0.2& -1.2& -1.7&&  689.2\cite{robin1988fluorination} \vspace{0.1cm}\\ 
\hline 
    \rule{0pt}{2.6ex}\textbf{MAD}&   13.2&   11.8&    4.8&    2.5 & &    2.2&    2.1&    1.2&    1.4&&\end{tabular}
\end{ruledtabular}

\end{table*}

\FloatBarrier

\begin{table*} 
\caption{Part I of the core IP benchmark table.}\label{IP1}
\begin{ruledtabular}
\begin{tabular}{l l c c c c p{0.01cm} c}
&& B3LYP & PBE0  & PBEh  & BHHLYP && Exp.\footnote{Values from Ref. ~\onlinecite{golze2020accurate} and references therein} \\ 
\hline 
     \rule{0pt}{2.6ex}CH\textsubscript{4}& C1s& -0.38& -0.11& +0.14& +0.05&&  290.84\vspace{0.15cm}\\ 
    C\textsubscript{2}H\textsubscript{6}& C1s& -0.52& -0.22& +0.04& -0.07&&  290.71\vspace{0.15cm}\\ 
    C\textsubscript{2}H\textsubscript{4}& C1s& -0.37& -0.06& +0.23& +0.12&&  290.82\vspace{0.15cm}\\ 
    C\textsubscript{2}H\textsubscript{2}& C1s& -0.45& -0.12& +0.17& +0.03&&  291.25\vspace{0.15cm}\\ 
    CO& C1s& +0.39& +0.62& +1.02& +1.01&&  296.23\vspace{0.15cm}\\ 
    CO& O1s& -2.17& -1.66& -1.48& -1.84&&  541.32\vspace{0.15cm}\\ 
    CO\textsubscript{2}& C1s& +0.10& +0.26& +0.87& +1.06&&  297.70\vspace{0.15cm}\\ 
    CO\textsubscript{2}& O1s& -3.41& -2.85& -2.51& -2.84&&  541.32\vspace{0.15cm}\\ 
    CF\textsubscript{4}& F1s& -3.68& -3.19& -2.93& -3.24&&  695.20\vspace{0.15cm}\\ 
    CF\textsubscript{4}& C1s& -0.14& -0.01& +0.61& +0.96&&  301.90\vspace{0.15cm}\\ 
    CH\textsubscript{3}F& C1s& -0.05& +0.18& +0.51& +0.49&&  293.56\vspace{0.15cm}\\ 
    CH\textsubscript{3}F& F1s& -3.63& -3.12& -3.01& -3.42&&  692.40\vspace{0.15cm}\\ 
    CHF\textsubscript{3}& C1s& +0.06& +0.22& +0.78& +0.92&&  299.16\vspace{0.15cm}\\ 
    CHF\textsubscript{3}& F1s& -3.46& -2.97& -2.76& -3.09&&  694.10\vspace{0.15cm}\\ 
    CH\textsubscript{3}OH& C1s& -0.05& +0.19& +0.49& +0.45&&  292.30\vspace{0.15cm}\\ 
    CH\textsubscript{3}OH& O1s& -2.89& -2.33& -2.13& -2.53&&  538.88\vspace{0.15cm}\\ 
    CH\textsubscript{2}O& C1s& +0.23& +0.45& +0.82& +0.82&&  294.38\vspace{0.15cm}\\ 
    CH\textsubscript{2}O& O1s& -3.03& -2.50& -2.30& -2.67&&  539.33\vspace{0.15cm}\\

\end{tabular}
\end{ruledtabular}
\end{table*}

\begin{table*} 
\caption{Part II of the core IP benchmark table.}\label{IP2}
\begin{ruledtabular}
\begin{tabular}{l l c c c c p{0.01cm} c}
&& B3LYP & PBE0  & PBEh  & BHHLYP && Exp.\footnote{Values from Ref. ~\onlinecite{golze2020accurate} and references therein} \\ 
\hline 
     \rule{0pt}{2.6ex}CH\textsubscript{3}OCH\textsubscript{3}& C1s& -0.19& +0.06& +0.35& +0.30&&  292.17\vspace{0.15cm}\\
     CH\textsubscript{3}OCH\textsubscript{3}& O1s& -2.99& -2.38& -2.17& -2.63&&  538.36\vspace{0.15cm}\\
     HCOOH& C1s& +0.04& +0.29& +0.76& +0.85&&  295.75\vspace{0.15cm}\\ 
     HCOOH& O1s(OH)& -3.11& -2.56& -2.28& -2.64&&  540.69\vspace{0.15cm}\\ 
    HCOOH& O1s(C=O)& -3.27& -2.74& -2.47& -2.81&&  539.02\vspace{0.15cm}\\ 
    acetone& O1s& -2.99& -2.45& -2.20& -2.57&&  537.73\vspace{0.15cm}\\ 
    acetone& C1s(C=O)& -0.82& -0.53& -0.07& -0.13&&  293.88\vspace{0.15cm}\\ 
    acetone& C1s(CH3)& -0.69& -0.40& -0.10& -0.18&&  291.23\vspace{0.15cm}\\ 
    HCO\textsubscript{2}CH\textsubscript{3}& O1s(C=O)& -2.98& -2.43& -2.15& -2.50&&  538.24\vspace{0.15cm}\\ 
    HCO\textsubscript{2}CH\textsubscript{3}& O1s(OCH3& -3.10& -2.50& -2.22& -2.64&&  539.64\vspace{0.15cm}\\ 
    CH\textsubscript{3}COOH& O1s(OH)& -3.12& -2.56& -2.27& -2.62&&  540.10\vspace{0.15cm}\\ 
    CH\textsubscript{3}COOH& O1s(C=O)& -3.28& -2.75& -2.47& -2.82&&  538.31\vspace{0.15cm}\\ 
    CH\textsubscript{3}COOH& C1s(CH3)& -0.49& -0.21& +0.10& +0.04&&  291.55\vspace{0.15cm}\\ 
    CH\textsubscript{3}COOH& C1s(COOH)& -0.38& -0.14& +0.39& +0.45&&  295.35\vspace{0.15cm}\\ 
    H\textsubscript{2}O& O1s& -2.95& -2.46& -2.28& -2.62&&  539.70\vspace{0.15cm}\\ 
    O\textsubscript{3}& O1s(terminal)& -3.83& -3.31& -2.75& -2.93&&  541.75\vspace{0.15cm}\\ 
    O\textsubscript{3}& O1s(middle)& -2.76& -2.14& -1.56& -1.85&&  546.44\vspace{0.15cm}\\ 
    N\textsubscript{2}& N1s& -1.72& -1.32& -0.96& -1.13&&  409.93\vspace{0.15cm}\\ 
    NH\textsubscript{3}& N1s& -1.67& -1.22& -0.98& -1.26&&  405.52\vspace{0.15cm}\\ 
    HCN& C1s& -0.44& -0.18& +0.13& +0.07&&  293.50\vspace{0.15cm}\\ 
    HCN& N1s& -2.01& -1.55& -1.22& -1.48&&  406.80\vspace{0.15cm}\\ 
    CH\textsubscript{3}CN& N1s& -2.09& -1.61& -1.25& -1.50&&  405.58\vspace{0.15cm}\\ 
    CH\textsubscript{3}CN& C1s(CH3)& -0.64& -0.35& -0.05& -0.13&&  292.88\vspace{0.15cm}\\ 
\end{tabular}
\end{ruledtabular}
\end{table*}

\begin{table*}
\caption{Part III of the core IP benchmark table.}\label{IP3}
\begin{ruledtabular}
\begin{tabular}{l l c c c c p{0.01cm} c}
&& B3LYP & PBE0  & PBEh  & BHHLYP && Exp.\footnote{Values from Ref. ~\onlinecite{golze2020accurate} and references therein} \\ 
\hline 
     \rule{0pt}{2.6ex}CH\textsubscript{3}CN& C1s(CN)& -0.70& -0.41& -0.06& -0.14&&  292.60\vspace{0.15cm}\\ 
     glycine& N1s& -1.95& -1.45& -1.16& -1.47&&  405.40\vspace{0.15cm}\\ 
    glycine& C1s(CH2)& -0.80& -0.52& -0.16& -0.22&&  292.30\vspace{0.15cm}\\ 
    glycine& C1s(COOH)& -0.60& -0.36& +0.19& +0.26&&  295.20\vspace{0.15cm}\\ 
    glycine& O1s(OH)& -3.40& -2.84& -2.54& -2.89&&  540.20\vspace{0.15cm}\\ 
    glycine& O1s(C=O)& -3.46& -2.92& -2.63& -2.97&&  538.40\vspace{0.15cm}\\ 
    pyridine& N1s& -2.42& -1.86& -1.52& -1.86&&  404.82\vspace{0.15cm}\\ 
    pyrrole& N1s& -2.18& -1.62& -1.33& -1.70&&  406.18\vspace{0.15cm}\\ 
    aniline& N1s& -1.91& -1.39& -1.12& -1.45&&  405.31\vspace{0.15cm}\\ 
    urea& C1s& -0.23& +0.02& +0.55& +0.63&&  294.84\vspace{0.15cm}\\ 
    urea& O1s& -3.21& -2.67& -2.39& -2.74&&  537.19\vspace{0.15cm}\\ 
    urea& N1s& -1.79& -1.31& -0.99& -1.26&&  406.09\vspace{0.15cm}\\ 
    methylamine& N1s& -1.78& -1.28& -1.03& -1.35&&  405.17\vspace{0.15cm}\\ 
    nitrobenzene& N1s& -2.00& -1.43& -0.74& -0.93&&  411.60\vspace{0.15cm}\\ 
    nitrobenzene& O1s& -3.29& -2.70& -2.29& -2.62&&  538.63\vspace{0.15cm}\\ 
    nitrobenzene& C1s(C1)& -1.33& -0.97& -0.53& -0.63&&  292.08\vspace{0.15cm}\\ 
    nitrobenzene& C1s(C2-4)& -1.05& -0.70& -0.30& -0.40&&  291.13\vspace{0.15cm}\\ 
    C\textsubscript{6}H\textsubscript{6}& C1s& -0.95& -1.13& -0.79& -0.39&&  290.38\vspace{0.15cm}\\ 
    phenylacetylene& C1s(C1)& +0.13& +0.49& +0.83& +0.68&&  289.75\vspace{0.15cm}\\ 
    phenylacetylene& C1s(C2)& -1.02& -0.65& -0.33& -0.50&&  290.55\vspace{0.15cm}\\ 
    phenylacetylene& C1s(C3)& -0.29& +0.09& +0.46& +0.32&&  289.75\vspace{0.15cm}\\ 
    phenylacetylene& C1s(C4-6)& -0.86& -0.49& -0.12& -0.26&&  290.16\vspace{0.15cm}\\ 
\hline 
    \rule{0pt}{2.6ex}\textbf{MAD} &&    1.68&    1.34&    1.21&    1.39&&\end{tabular}
\end{ruledtabular}
\end{table*}

\FloatBarrier

\begin{figure*}[tb]
    \centering
    \includegraphics[width=1.0\textwidth]{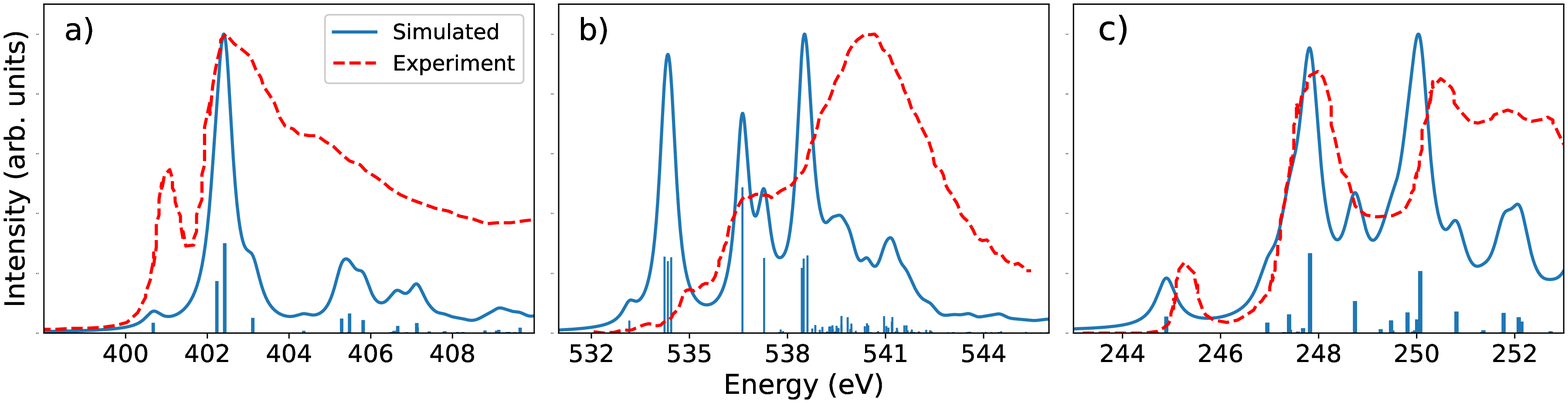}
    \caption{GW2X* corrected LR-TDDFT spectra of a) solid ammonia N K-edge, b) ice 1h O K-edge and c) solid argon Ar L\textsubscript{2,3}-edge in periodic boundary conditions. The GW2X* corrections amount to blue shifts of 3.3 eV, 2.3 eV and 3.2 eV for ammonia, ice and argon, respectively. A Lorentzian broadening of fwhm 0.5 eV was applied and calculated intensities uniformly scaled to match experiments.}
    \label{fig:solids}
\end{figure*}

\section{GW2X* in extended systems}

Figure \ref{fig:solids} shows the results of the simplified GW2X* method applied to extended systems in periodic boundary conditions. The GW2X* shifts are 3.3 eV, 2.3 eV and 3.2 eV for solid ammonia, ice 1h and solid argon, respectively. They differ from the original GW2X shifts by 0.4 eV, 0.5 eV and 0.2 eV. While these differences are not critical, they certainly affect the alignment of the first peak of each spectrum. Apart from the use of approximate GW2X*, all computational details for these calculations are the same as those using standard GW2X in the main paper.

\FloatBarrier

\nocite{*}
\bibliography{supp}